\documentclass[useAMS,usenatbib]{mnras}
\usepackage{natbib}
\usepackage{amsmath}
\usepackage{url}
\usepackage{longtable}
\usepackage{aas_macros}
\usepackage{amssymb}
\usepackage{graphicx}
\usepackage{deluxetable}
\usepackage{pdflscape}
\usepackage{multirow}
\usepackage{color}

\newcommand{\about}{$\sim\!\!$~}
\newcommand{\kms}{\,km\,s$^{-1}$}
\newcommand{\dm}{$\Delta m_{15} (B)$}

\def\lsim{\hbox{\rlap{\raise 0.425ex\hbox{$<$}}\lower 0.65ex\hbox{$\sim$}}}
\def\gsim{\hbox{\rlap{\raise 0.425ex\hbox{$>$}}\lower 0.65ex\hbox{$\sim$}}}

\def\arcsec{\hbox{$^{\prime\prime}$}}

\title[UV Diversity of SNe~Ia]{Ultraviolet Diversity of Type Ia Supernovae}

\def\illast{1}
\def\illphys{2}
\def\tamu{3}
\def\berk{4}
\def\stsci{5}
\def\mpia{6}
\def\cfa{7}
\def\moore{8}
\def\ut{9}
\def\az{10}
\def\ab{11}
\def\mil{12}
\def\aar{13}

\begin{document}

\author[Foley, et~al.]{Ryan~J.~Foley$^{\illast,\illphys}$\thanks{E-mail:rfoley@illinois.edu},
Yen-Chen~Pan$^{\illast}$,
P.~Brown$^{\tamu}$,
A.~V.~Filippenko$^{\berk}$,
O.~D.~Fox$^{\stsci}$,
\newauthor
W.~Hillebrandt$^{\mpia}$,
R.~P.~Kirshner$^{\cfa,\moore}$,
G.~H.~Marion$^{\ut}$,
P.~A.~Milne$^{\az}$,
J.~T.~Parrent$^{\cfa}$,
\newauthor
G.~Pignata$^{\ab,\mil}$,
M.~D.~Stritzinger$^{\aar}$
\\
$^{\illast}$Astronomy Department, University of Illinois at Urbana-Champaign, 1002 W.\ Green Street, Urbana, IL 61801, USA\\
$^{\illphys}$Department of Physics, University of Illinois at Urbana-Champaign, 1110 W.\ Green Street, Urbana, IL 61801, USA\\
$^{\tamu}$George P.\ and Cynthia Woods Mitchell Institute for Fundamental Physics \& Astronomy, Texas A.\ \& M.\ University,\\ Department of Physics and Astronomy, 4242 TAMU, College Station, TX 77843, USA\\
$^{\berk}$Department of Astronomy, University of California, Berkeley, CA 94720-3411, USA\\
$^{\stsci}$Space Telescope Science Institute, 3700 San Martin Drive, Baltimore, MD 21218, USA\\
$^{\mpia}$Max-Planck-Institut f\"{u}r Astrophysik, Karl-Schwarzschild-Strasse 1, D-85748 Garching bei M\"{u}nchen, Germany\\
$^{\cfa}$Harvard-Smithsonian Center for Astrophysics, 60 Garden Street, Cambridge, MA 02138, USA\\
$^{\moore}$Gordon and Betty Moore Foundation, 1661 Page Mill Road, Palo Alto CA 94304, USA\\
$^{\ut}$University of Texas at Austin, 1 University Station C1400, Austin, TX, 78712-0259, USA\\
$^{\az}$University of Arizona, Steward Observatory, 933 N. Cherry Ave., Tucson, AZ 85719, USA\\
$^{\ab}$Departamento de Ciencias Fisicas, Universidad Andres Bello, Avda.\ Republica 252, Santiago, Chile\\
$^{\mil}$Millennium Institute of Astrophysics, Avda.\ Republica 252, Santiago, Chile\\
$^{\aar}$Department of Physics and Astronomy, Aarhus University, Ny Munkegade 120, DK-8000 Aarhus C, Denmark
}
\date{Accepted  . Received   ; in original form  }
\pagerange{\pageref{firstpage}--\pageref{lastpage}} \pubyear{2016}
\maketitle
\label{firstpage}

\begin{abstract}
  Ultraviolet (UV) observations of Type Ia supernovae (SNe~Ia) probe
  the outermost layers of the explosion, and UV spectra of SNe~Ia are
  expected to be extremely sensitive to differences in progenitor
  composition and the details of the explosion.  Here we present the
  first study of a sample of high signal-to-noise ratio SN~Ia spectra
  that extend blueward of 2900~\AA.  We focus on spectra taken within
  5~days of maximum brightness.  Our sample of ten SNe~Ia spans the
  majority of the parameter space of SN~Ia optical diversity.  We find
  that SNe~Ia have significantly more diversity in the UV than in the
  optical, with the spectral variance continuing to increase with
  decreasing wavelengths until at least 1800~\AA\ (the limit of our
  data).  The majority of the UV variance correlates with optical
  light-curve shape, while there are no obvious and unique
  correlations between spectral shape and either ejecta velocity or
  host-galaxy morphology.  Using light-curve shape as the primary
  variable, we create a UV spectral model for SNe~Ia at peak
  brightness.  With the model, we can examine how individual SNe vary
  relative to expectations based on only their light-curve shape.
  Doing this, we confirm an excess of flux for SN~2011fe at short
  wavelengths, consistent with its progenitor having a subsolar
  metallicity.  While most other SNe~Ia do not show large deviations
  from the model, ASASSN-14lp has a deficit of flux at short
  wavelengths, suggesting that its progenitor was relatively metal
  rich.
\end{abstract}



\section{Introduction}\label{s:intro}

Type Ia supernovae (SNe~Ia) are thermonuclear explosions of C/O white
dwarfs in binary systems \citep[see][for a review]{Hillebrandt00}.
Variable amounts of $^{56}$Ni are produced in the explosion, which
dictates the peak luminosity and photospheric temperature of the SN;
this in turn determines the light-curve shape \citep{Kasen07:wlr},
which has been empirically measured \citep{Phillips93}.  These
relatively simple physical connections make SNe~Ia particularly good
distance indicators, providing the first direct evidence for the
currently accelerating expansion of the Universe
\citep{Riess98:Lambda, Perlmutter99} and continuing to provide
critical constraints on cosmological parameters
\citep[e.g.,][]{Betoule14, Rest14}.

Despite our general physical understanding of SNe~Ia, there remain
significant questions about their progenitor systems (e.g., what is
the nature of the companion star) and the explosion mechanism (e.g.,
how is the flame ignited and how does it propagate).  Lacking this
fundamental knowledge hinders future theoretical and observational
investigations.

A unique way to probe the physics of SN~Ia explosions and progenitors
is through their ultraviolet (UV) spectra.  While the main source of
opacity for a SN atmosphere near peak brightness is electron
scattering at most optical wavelengths, the UV opacity is dominated by
a forest of overlapping lines from Fe-group elements
\citep[e.g.,][]{Baron96, Pinto00}.  UV photons are repeatedly absorbed
and re-emitted, and typically are scattered redward where they
eventually escape the expanding SN ejecta.  Therefore, the UV is
essential for understanding the optical emission of a SN~Ia
\citep{Sauer08} as well as being extremely sensitive to both the
progenitor composition and explosion mechanism.  Because of the high
opacities in the UV, we can use UV spectroscopy to directly probe the
composition of the outermost layers of the SN which are transparent at
optical wavelengths soon after explosion.

After correcting for light-curve shape, SN~Ia luminosity still depends
significantly on host-galaxy environment (\citealt{Kelly10};
\citealt{Lampeitl10:host}; \citealt{Sullivan10}; however, for an
alternative explanation, see \citealt{Kim14}).  This may indicate that
environmental effects or progenitor properties affect our luminosity
calibration.  The UV can potentially improve our physical
understanding of the relationship between host-galaxy mass and SN~Ia
luminosity.

In particular, progenitor metallicity should affect the amount of
radioactive material generated in the explosion \citep{Timmes03} and
the relationship between SN luminosity and light-curve shape
\citep{Mazzali01, Mazzali06, Podsiadlowski06}, while also shaping the
UV spectrum \citep[e.g.,][]{Hoflich98, Lentz01}.  Such a relation may
significantly impact the overall SN calibration and increase the
scatter in SN distance measurements \citep{Foley13:met}.
Additionally, if the mean SN progenitor metallicity has evolved with
cosmic time, we would expect a systematic bias in cosmological
distance estimates.

To address these questions, we have undertaken a major program to
obtain UV spectra of SNe~Ia with the {\it Hubble Space Telescope}
({\it HST}).  Until now, we have focused on the detailed study of
individual events \citep{Kirshner93, Foley12:11iv, Foley13:met,
  Foley14:14j, Foley13:ca, Pan15:13dy}.  And while other studies have
examined samples of SNe~Ia with UV spectra \citep{Foley08:uv, Cooke11,
  Maguire12, Wang12:uv}, those data either did not extend blueward of
\about 2900~\AA\ or had low signal-to-noise ratio (S/N).

Here we present the first study of a sample of SNe~Ia with
near-maximum-light space-UV (extending to $<2000$~\AA) spectra.  With
our sample, we are able to investigate how the spectra are influenced
by other properties of the SN.  With these initial results, we can
account for correlations between spectral features and light-curve
shape, which, in turn, can isolate effects related to other physical
properties such as progenitor metallicity.

This paper is structured as follows.  We present new observations of
two SNe~Ia and add those data to our previous sample in
Section~\ref{s:obs}.  The characteristics of the sample are examined
and the spectral properties are analyzed in Section~\ref{s:anal}.  We
discuss our results and conclude in Section~\ref{s:disc}.


\section{Data \& Observations}\label{s:obs}

ASASSN-14lp was discovered in NGC~4666 on 2014 December 9.6 \citep[all
dates herein are UT;][]{Holoien14:14lp}.  Spectroscopic observations
on 2014 December 10.8 indicated that it was a young SN~Ia
\citep{Thorstensen14}.  SN~2015F in NGC~2442 was discovered on 2015
March 9.8 \citep{Monard15} and spectra taken on 2015 March 11.0 showed
it was a young SN~Ia \citep{Fraser15}.  As part of our ongoing program
to obtain UV spectra of SNe~Ia, we triggered {\it HST} to observe
ASASSN-14lp and SN~2015F in Cycle 22 \citep{Foley14:14lp,
  Foley15:15f}.  ASASSN-14lp and SN~2015F peaked in $B$ on 2014
December 24.25 \citep{Shappee15} and 2015 March 24.98 \citep{Im15},
respectively.

ASASSN-14lp and SN~2015F were observed by \textit{HST} using the STIS
spectrograph on 2014 December 19.81 and 2015 March 22.66,
corresponding to $t = -4.4$ and $-2.3$~days relative to $B$ maximum,
respectively.  The observations were obtained over one orbit per SN
with three different gratings, all with the $52\arcsec \times
0.\farcs2$ slit.  Exposures of 1393~s (for ASASSN-14lp; 1345~s for
SN~2015F) utilized the near-UV MAMA detector and the G230L grating.
For each SN, exposures of 100~s were taken with both the CCD/G430L and
CCD/G750L setups.  The three setups yield a combined wavelength range
of 1615--10,230~\AA.

The data were reduced using the standard \textit{HST} Space Telescope
Science Data Analysis System (STSDAS) routines to bias-subtract,
flat-field, extract, wavelength-calibrate, and flux-calibrate each SN
spectrum.  Similar reductions were performed for the other spectra
used in this study \citep{Foley12:11iv, Foley13:met, Foley14:14j,
  Foley13:ca, Pan15:13dy}.

In addition to the new data, we include near-maximum-light ($|t|
\lesssim 5$~days) {\it HST} spectra of SNe~1992A \citep{Kirshner93},
2011by \citep{Foley13:met}, 2011fe \citep{Foley13:ca}, 2011iv
\citep{Foley12:11iv}, 2013dy \citep{Pan15:13dy}, and 2014J
\citep{Foley14:14j}.  We present the maximum-light {\it HST} spectrum
of SN~2012cg, originally published by \citet{Amanullah15}, but
rereduced as described above (with the modest difference that the UV
spectrum was obtained with the STIS CCD/G230LB setup and there was no
CCD/G750L observation).  To this, we also add the high-S/N
near-maximum-light {\it Swift} UV spectrum of SN~2009ig
\citep{Foley12:09ig}.  While the quality of the SN~2009ig spectrum is
not as high as that of the {\it HST} spectra and it does not probe
shortward of \about 2500~\AA, it still provides useful information in
the near UV.

We correct all spectra for both Milky Way reddening \citep{Schlegel98,
  Schlafly11} and host-galaxy reddening as derived from the SN light
curves \citep{Phillips99, Foley12:09ig, Foley12:11iv, Foley14:14j,
  Foley13:met, Pereira13, Silverman13, Amanullah14, Im15, Pan15:13dy,
  Shappee15}.

Our total sample contains 10 SNe.  Of these, six (SNe~2011fe, 2011iv,
2013dy, 2014J, 2015F, and ASASSN-14lp) have {\it HST} UV spectral
sequences consisting of two or more epochs of UV spectroscopy.  While
SN~2014J has an excellent spectral series, it also suffers from
significant host-galaxy reddening \citep[e.g.,][]{Foley14:14j,
  Goobar14, Brown15}, reducing the S/N in the far-UV and making the
intrinsic continuum difficult to infer.


\section{Analysis}\label{s:anal}

\subsection{Sample Demographics}

The diversity of peak-luminosity optical SN~Ia spectra is primarily
driven by the photospheric temperature, which changes the ionization
state of elements, and the ejecta velocity, which shifts and broadens
absorption features.  This diversity can be parameterized by the
pseudoequivalent widths (pEWs) of the \ion{Si}{II} $\lambda 5972$ and
\ion{Si}{II} $\lambda 6355$ features \citep{Branch06}.  Alternatively,
the diversity can be described by light-curve shape (e.g., \dm) and
the maximum-light velocity of \ion{Si}{II} $\lambda 6355$, $v_{\rm
  Si~II}^{0}$ \citep{Wang09:2pop}.  Photospheric temperature is highly
correlated with both \dm\ and the relative strengths of \ion{Si}{II}
$\lambda\lambda 5972$, 6355 \citep{Nugent02}; also, the pEW of
\ion{Si}{II} $\lambda 6355$ is correlated with $v_{\rm Si~II}^{0}$
\citep{Foley11:vel}.

\begin{figure}
\begin{center}
\includegraphics[angle=0,width=3.2in]{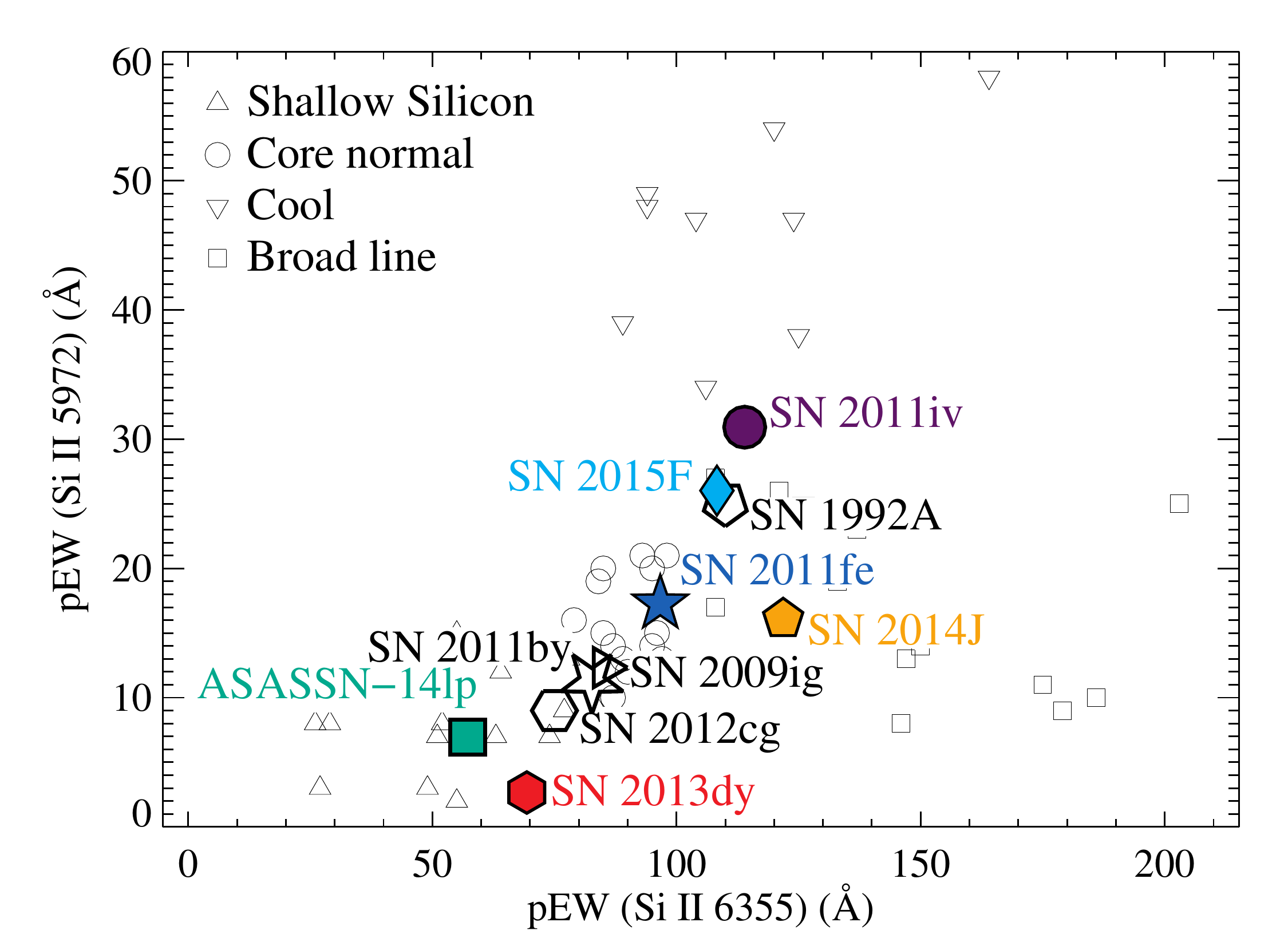}
\caption{pEWs of the \ion{Si}{II} $\lambda 5972$ and \ion{Si}{II}
  $\lambda 6355$ lines for a sample of SNe~Ia \citep{Branch06}, with
  symbol shapes corresponding to the designations defined by
  \citet{Branch06}.  The large, labeled points indicate the sample of
  SNe~Ia with near-maximum-brightness high-S/N UV spectra.  The
  coloured points correspond to the SNe with {\it HST} spectral
  sequences.}\label{f:subclass}
\end{center}
\end{figure}

Figure~\ref{f:subclass} displays the \ion{Si}{II} parameter space for
a sample of SNe~Ia \citep{Branch06}.  \citet{Branch06} subclassified
SNe~Ia by these measurements.  There is a general trend from weak
lines (small pEW values) to strong lines.  The SNe with the weakest
lines are called ``Shallow Silicon'' and have spectra (and light
curves) similar to those of SN~1991T \citep{Filippenko92:91T,
  Phillips92}.  The SNe with slightly stronger lines are called ``Core
Normal.''  The SNe with the strongest \ion{Si}{II} $\lambda 5972$
lines share properties with SN~1991bg \citep{Filippenko92:91bg,
  Leibundgut93} and are called ``Cool.''  Finally, those SNe with
particularly strong \ion{Si}{II} $\lambda 6355$, which generally
correlates with high ejecta velocities, are called ``Broad Line.''

Figure~\ref{f:subclass} also displays where the SNe~Ia with
near-maximum-light UV spectra fall in this parameter space.  For most
of the sample, we were able to measure the pEWs directly from the {\it
  HST} spectra.  However, we used other optical spectra to determine
these values for SNe~1992A, 2009ig, 2011by, and 2012cg
\citep{Kirshner93, Foley12:09ig, Silverman12:12cg, Silverman13}.

The ten SNe~Ia with high-S/N near-maximum-light spectra span most of
the above parameter space.  While the SNe generally include most of
the Shallow Silicon and Core Normal region, they only barely reach the
Cool and Broad-Line subclasses.  Moreover, SN~2014J, which has very
high and uncertain reddening, is the only true Broad-Line SN.  Because
of the uncertain reddening, we remove SN~2014J from our subsequent
analysis, making our final sample primarily a continuum from
SN~1991T-like SNe to cool, but not extremely cool SNe (i.e., similar
to SN~1986G; \citealt{Phillips87}), with no true Broad-Line SNe.

Optical spectral properties also correlate with light-curve shape
\citep{Nugent95} and host-galaxy morphology \citep[e.g.,][]{Hamuy00,
  Howell01}.  We present basic parameters, including light-curve
shape, velocity, and host-galaxy morphology for the sample in
Table~\ref{t:params}.  Light-curve shape measurements were taken from
the literature \citep{Phillips99, Foley12:09ig, Foley12:11iv,
  Foley14:14j, Foley13:met, Pereira13, Silverman13, Amanullah14, Im15,
  Pan15:13dy, Shappee15}.  For the ejecta velocity, we measure the
velocity of the \ion{Si}{II} absorption-line minimum in the optical
spectra, and correct those data to their maximum-light values
\citep{Foley11:vgrad}, $v_{\rm Si~II}^{0}$.  Host-galaxy morphology
measurements were obtained from the NASA/IPAC Extragalactic Database
(NED).

\begin{deluxetable}{llll}
\tabletypesize{\footnotesize}
\tablewidth{0pt}
\tablecaption{UV SN Sample Properties\label{t:params}}
\tablehead{
\colhead{SN} & \colhead{$\Delta m_{15} (B)$} & \colhead{$v_{\rm Si~II}^{0}$} & \colhead{Host}\\
\colhead{} & \colhead{(mag)} & \colhead{(\kms)} & \colhead{Morphology}}

\startdata

1992A       & 1.47 & $-14$,000 & S0\\
2009ig      & 0.89 & $-13$,500 & Sa\\
2011by      & 1.14 & $-10$,300 & Sbc\\
2011fe      & 1.10 & $-10$,400 & Scd\\
2011iv      & 1.69 & $-10$,400 & E1\\
2012cg      & 0.86 & $-11$,000 & Sa\\
2013dy      & 0.92 & $-10$,400 & Sdm\\
2014J       & 0.95 & $-11$,900 & Sm\\
ASASSN-14lp & 0.80 & $-11$,100 & Sc\\
2015F       & 1.26 & $-10$,300 & Sbc

\enddata

\vspace{-0.6cm}

\tablecomments{\dm\ measurements have a typical uncertainty of
  0.03~mag.  $v_{\rm Si~II}^{0}$ measurements have a typical
  uncertainty of 250~\kms.\vspace{-0.4cm}}
\tablerefs{\citet{Phillips99, Foley12:09ig, Foley12:11iv, Foley14:14j,
    Foley13:met, Pereira13, Silverman13, Amanullah14, Im15,
    Pan15:13dy, Shappee15}}

\end{deluxetable}

\subsection{Spectral Correlations}

To directly compare the UV spectral properties of our sample, we
generate a smoothed spectrum of each SN using an inverse-variance
Gaussian filter \citep{Blondin06} and scale the spectra to have
roughly the same flux at 4000~\AA.  The SNe all have relatively
similar optical spectra, and therefore the exact choice of scaling (or
wavelength region where the scaling occurs) does not significantly
affect any results.

\begin{figure*}
\begin{center}
\includegraphics[angle=0,width=6.4in]{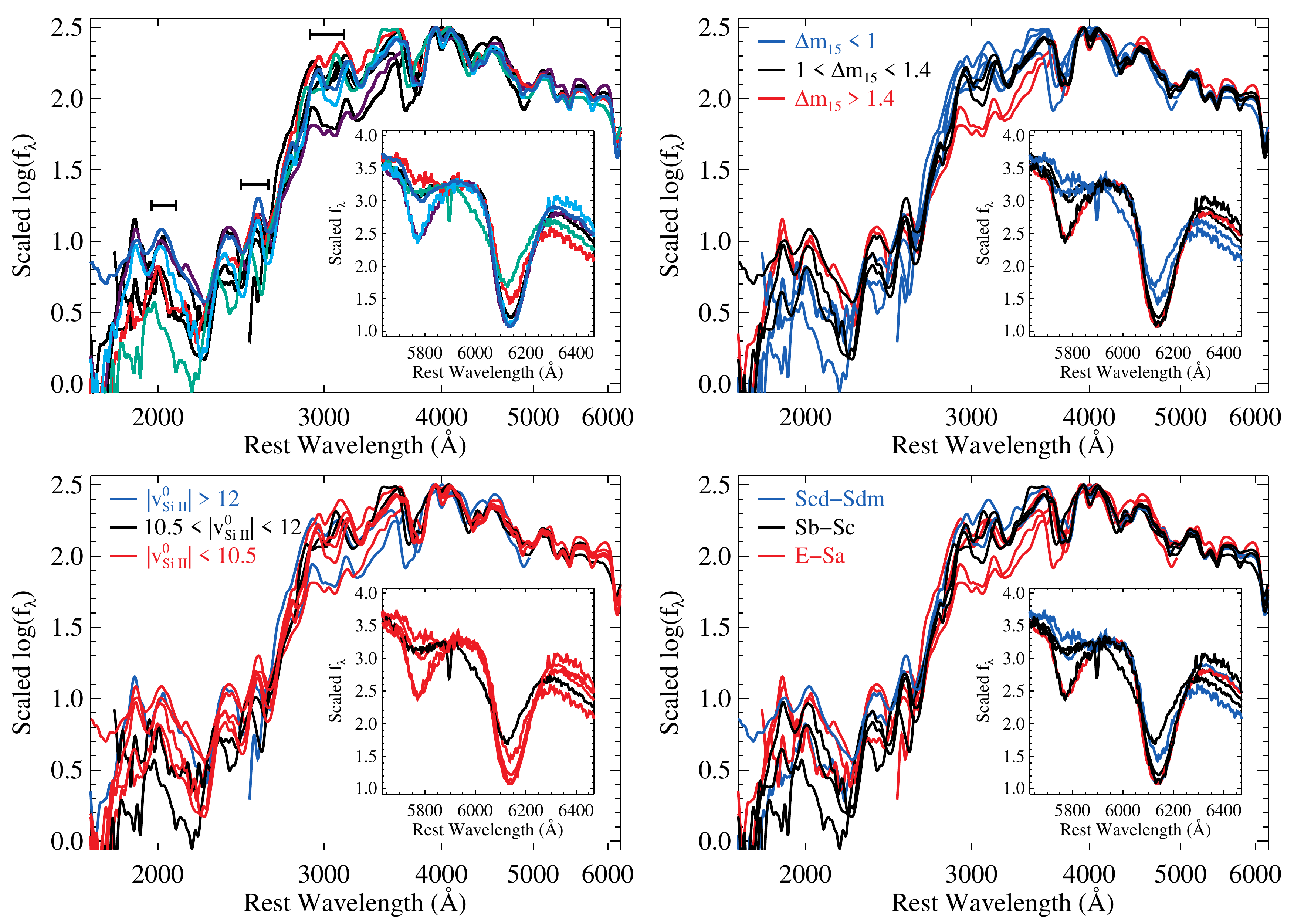}
\caption{Near-maximum-light UV-optical spectra of several SNe~Ia.  In
  the main panels, the spectra have been smoothed using an
  inverse-variance Gaussian filter \citep{Blondin06} and scaled to
  have a similar flux level at \about 4000~\AA.  In the subpanels, the
  region near \ion{Si}{II} $\lambda\lambda 5972$, 6355 is shown in
  detail (with different flux scaling).  The top-left panel displays
  the spectra with colours corresponding to those in
  Figure~\ref{f:subclass}.  The marked regions indicate, from shorter
  to longer wavelengths, the $f_{3025}$, $f_{2535}$, and $f_{2030}$
  flux regions.  The other panels display the same spectra, but
  coloured by different properties, with the top-right, bottom-left,
  and bottom-right panels representing light-curve shape, ejecta
  velocity, and host-galaxy morphology, respectively.  The blue,
  black, and red curves correspond to (respectively) $\Delta m_{15}
  (B) < 1$, $1 < \Delta m_{15} (B) < 1.4$, and $\Delta m_{15} (B) >
  1.4$~mag; $|v_{\rm Si~II}^{0}| > 12$,000, 10,$500 < |v_{\rm
    Si~II}^{0}| < 12$,000, and $|v_{\rm Si~II}^{0}| < 10$,500~\kms;
  and E through Sa, Sb through Sc, and Scd through Sdm.}\label{f:spec}
\end{center}
\end{figure*}

We present these UV through optical spectra in Figure~\ref{f:spec},
including a zoomed-in region near \ion{Si}{II} $\lambda\lambda 5972$,
6355.  The spectra are relatively similar over the range
4000--6000~\AA, but vary significantly for $\lambda < 3500$~\AA.
Increased diversity at shorter wavelengths has been shown for other
SN~Ia samples \citep{Ellis08, Foley08:comp, Foley12:sdss, Maguire12},
but no previous study has examined the region below \about 2500~\AA.

The subpanels of Figure~\ref{f:spec} display the spectra coloured by
their light-curve shape, ejecta velocity, and host-galaxy morphology.
Examining the spectra, there is a trend between light-curve shape and
the flux level at \about 3000~\AA\ (relative to \about 4000~\AA), with
faster decliners having lower flux.  This is similar to the trend seen
between the UV ratio and light-curve shape \citep{Foley08:uv}, but
with a different normalization wavelength.  The flux at other
wavelengths is also correlated with light-curve shape, but not as
strongly as at \about 3000~\AA.

There are no clear trends between spectral flux levels and ejecta
velocity or host-galaxy morphology.  However, the velocity of UV
features (unsurprisingly) correlates with $v_{\rm Si~II}^{0}$.

\begin{figure*}
\begin{center}
\includegraphics[angle=0,width=6.4in]{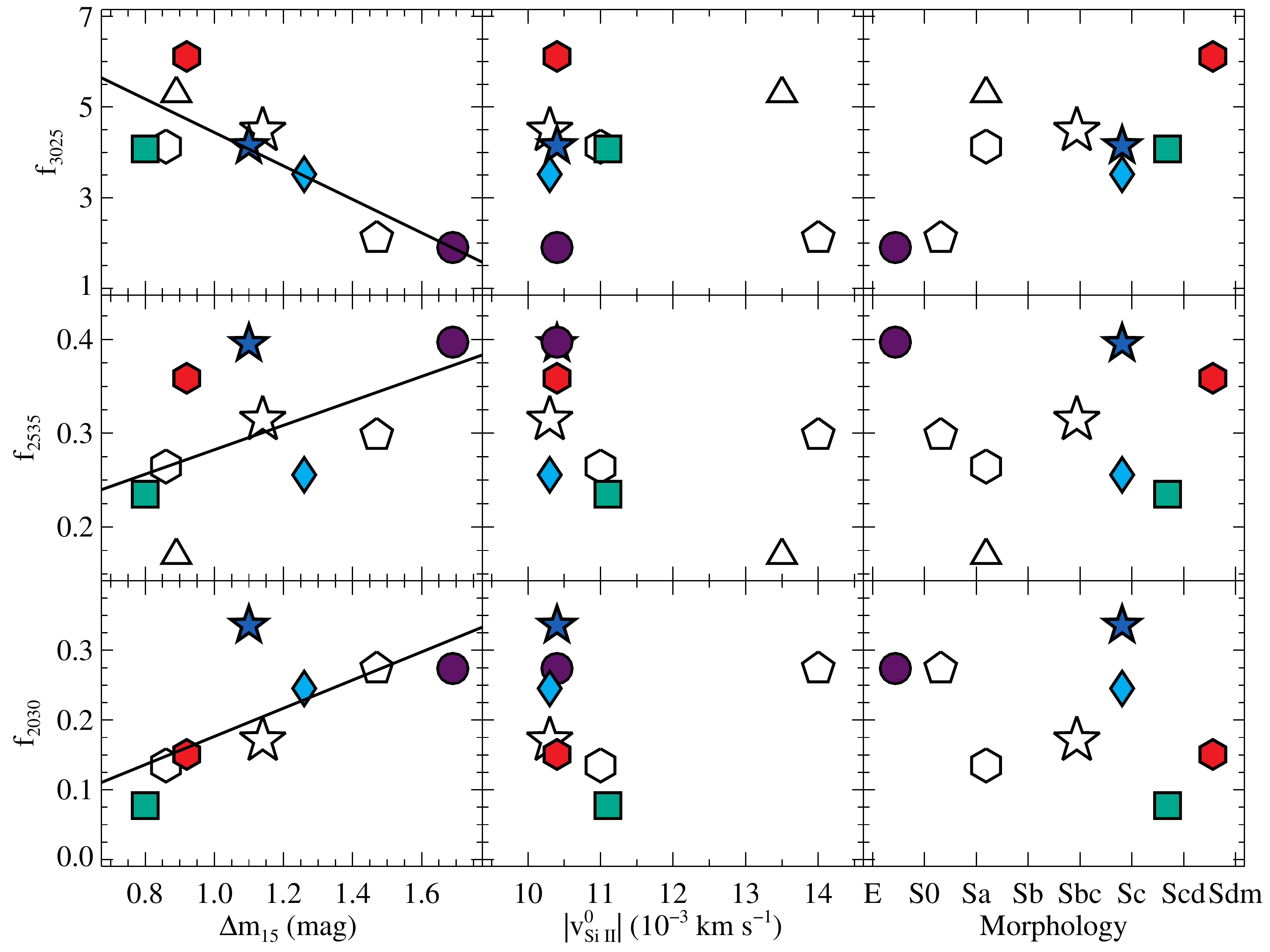}
\caption{Flux measurements (relative to 10 times the flux at 4000~\AA)
  for 1970--2090 ($f_{2030}$; bottom row), 2450--2620 ($f_{2535}$;
  middle row), and 2900--3150~\AA\ ($f_{3025}$; top row) as a function
  of \dm\ (left column), $v_{\rm Si~II}^{0}$ (middle column), and
  host-galaxy morphology (right column).  Symbols and colours
  correspond to objects as labeled in Figure~\ref{f:subclass}.  In the
  left column, we show the best-fitting line between \dm\ and the flux
  values.}\label{f:rat}
\end{center}
\end{figure*}

We define three regions of interest, corresponding to wavelength
ranges of 1970--2090, 2450--2620, and 2900--3150~\AA.  These represent
the far-UV region that is theoretically affected most by progenitor
metallicity \citep[e.g.,][]{Lentz00}, a mid-UV region on top of a
high-variation feature, and the near-UV feature that clearly and
strongly correlates with light-curve shape.  We label the median flux
in these regions, relative to 10 times that of the peak flux near
4000~\AA, as $f_{2030}$, $f_{2535}$, and $f_{3025}$, respectively.  We
display these flux values as a function of SN properties in
Figure~\ref{f:rat}.

The flux in each of these regions is correlated with light-curve
shape, although to varying degrees (Pearson correlation coefficient of
$r = 0.71$, $0.52$, and $-0.82$, respectively).  With the exception of
$f_{3025}$, none is highly correlated with ejecta velocity or
host-galaxy morphology.  For $f_{3025}$, it is correlated with
host-galaxy morphology ($r = 0.66$), but this is likely because
light-curve shape and host-galaxy morphology are highly correlated
\citep[e.g.,][]{Hamuy00, Howell01}.

To determine if there were additional correlations beyond light-curve
shape, we first fit linear trends between the various flux
measurements and \dm.  Examining the residuals, we find that there is
a somewhat strong correlation between ejecta velocity and $f_{2535}$
($r = -0.63$), but the lack of many high-velocity SNe prevents a
robust conclusion about a physical connection.

\subsection{Spectral Model}

Despite the strong correlations between light-curve shape and flux,
there is additional spectral diversity that is not described by this
single parameter.  This is obvious when comparing SNe~2011by and
2011fe, which have similar light-curve shapes but different UV
continua \citep{Foley13:met, Graham15}.  To assess how much an
individual spectrum deviates from a single parameterization, we
generated a data-driven model of the UV spectra.  Here, we fit the
smoothed flux for all spectra in our sample at each wavelength as a
function of \dm\ such that
\begin{equation}
  f_{\lambda} = f_{1.1, \lambda} + s_{\lambda} \times (\Delta m_{15} (B) - 1.1), \label{e:model}
\end{equation}
where $f_{1.1, \lambda}$ represents the spectrum of a nominal $\Delta
m_{15} (B) = 1.1$~mag SN~Ia and $s_{\lambda}$ is the deviation from
that spectrum for a hypothetical $\Delta m_{15} (B) = 2.1$~mag SN~Ia.
We present these parameters in Table~\ref{t:fitspec}.

\begin{deluxetable}{crr}
\tabletypesize{\footnotesize}
\tablewidth{0pt}
\tablecaption{UV Spectral Model Parameters\label{t:fitspec}}
\tablehead{
\colhead{Wavelength} & \colhead{$f_{1.1, \lambda}$} & \colhead{$s_{\lambda}$}\\
\colhead{(\AA)} & \colhead{} & \colhead{}}

\startdata

1700 &   0.0080  & 0.098\\
1705 & $-0.0010$ & 0.090\\
1710 &   0.0090  & 0.041\\
1715 &   0.0170  & 0.029\\
1720 &   0.0270  & 0.007\\
1725 &   0.0060  & 0.050\\
1730 &   0.0000  & 0.056\\
1735 &   0.0380  & 0.014\\
1740 &   0.0160  & 0.027\\
1745 &   0.0430  & 0.017

\enddata

\vspace{-0.6cm}
\tablecomments{Table~\ref{t:fitspec} is published in its entirety in
  the electronic edition of {\it Monthly Notices of the Royal
    Astronomical Society}.  A portion is shown here for guidance
  regarding its form and content.}

\end{deluxetable}

\begin{figure}
\begin{center}
\includegraphics[angle=0,width=3.2in]{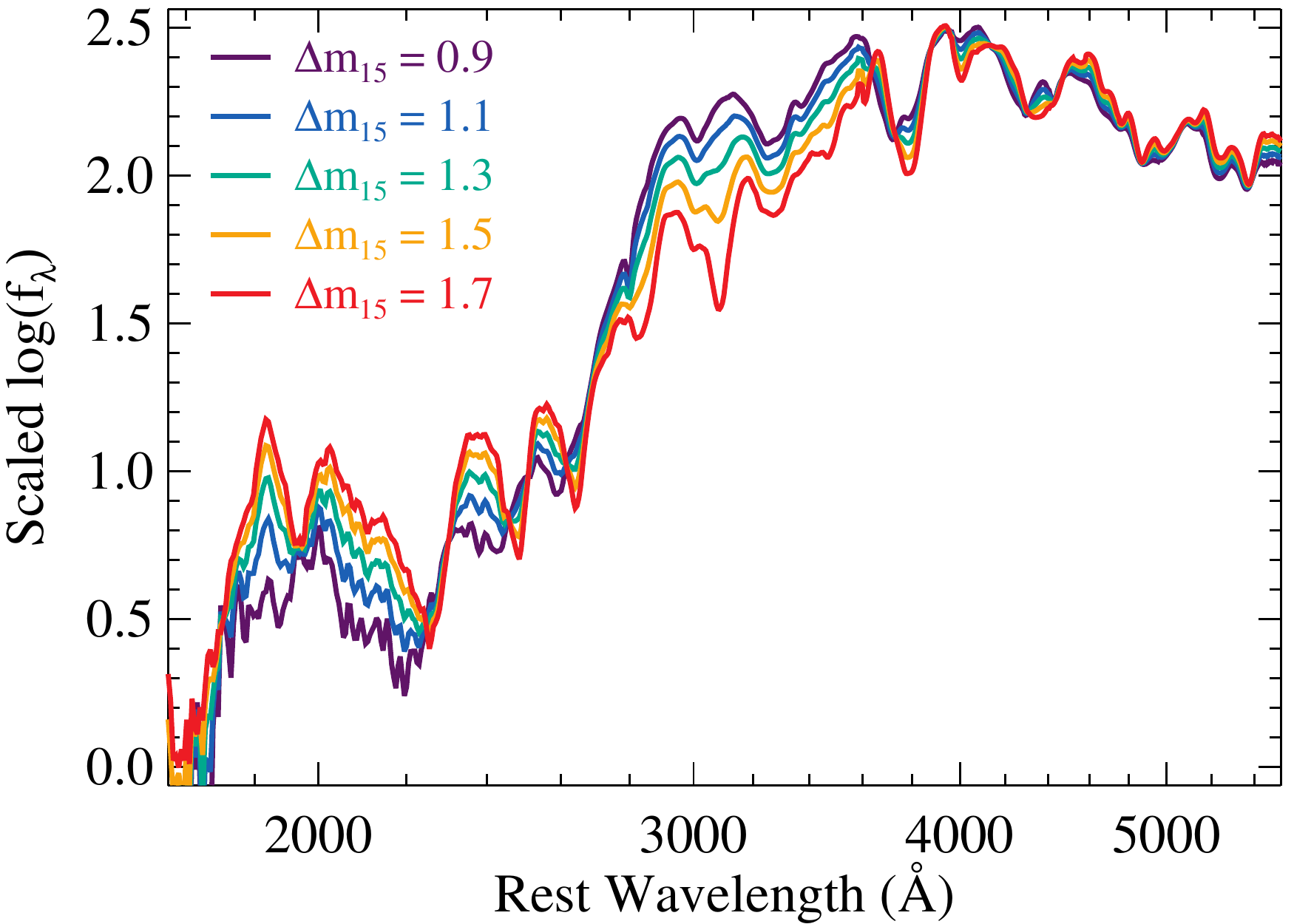}
\caption{Model near-maximum-light UV SN~Ia spectra as described by
  Equation~\ref{e:model}.  The purple, blue, green, gold, and red
  curves represent models with $\Delta m_{15} (B) = 0.9$, 1.1, 1.3,
  1.5, and 1.7~mag, respectively.}\label{f:model_example}
\end{center}
\end{figure}

Figure~\ref{f:model_example} displays model spectra for several values
of \dm.  The model spectra appear similar to the data.  Additionally,
the general trends observed in the data (e.g., the correlation between
light-curve shape and the flux at \about 3000~\AA) exist in the model.
There are other obvious trends in the model that are harder to
directly visualize in the spectra, such as increasing flux at \about
2000~\AA\ with increasing \dm\ --- however, these trends are present
in the data (see Figure~\ref{f:rat}).

The models have a pivot point at \about 2700~\AA.  This implies that
all SNe~Ia should have the same flux at \about 2700~\AA\ relative to
their optical flux (and in particular the flux at \about 4000~\AA).
Blueward of 2700~\AA, faster decliners have relatively higher flux,
while slower decliners have relatively more flux in the region \about
2700--4000~\AA.

\begin{figure*}
\begin{center}
\includegraphics[angle=0,width=6.4in]{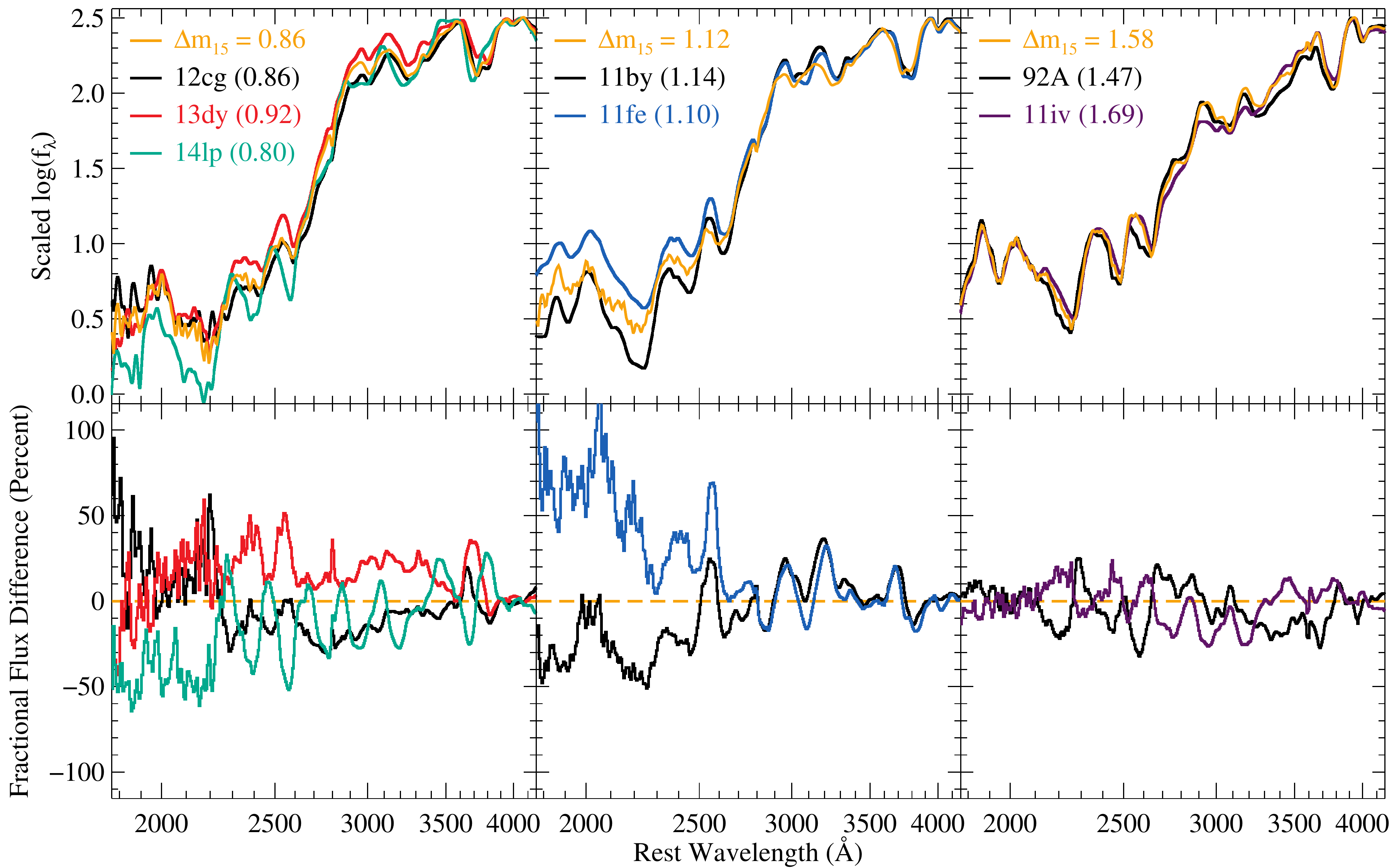}
\caption{(Top panels): Smoothed near-maximum-light UV spectra of
  several SNe~Ia.  The left, middle, and right panels present spectra
  of (respectively) SN~2012cg, SN~2013dy, and ASASSN-14lp; SNe~2011by
  and 2011fe; and SNe~1992A and 2011iv.  The spectra are separated
  such that each panel has spectra of SNe having similar light-curve
  shapes, where each \dm\ is labeled in parentheses.  The spectrum
  from a given SN is coloured to match that presented in
  Figure~\ref{f:subclass}.  The gold curves represent the model
  spectra, given by Equation~\ref{e:model}, for a nominal SN with a
  light-curve shape that is similar to those of the SNe whose spectra
  are displayed in that panel: $\Delta m_{15} (B) = 0.86$, 1.12, and
  1.58~mag for the left, middle, and right panels, respectively.
  (Bottom panels): Fractional flux differences from the model spectra
  presented in the panels above.  By dividing by the model spectrum,
  one can directly compare differences between
  SNe.}\label{f:spec_model}
\end{center}
\end{figure*}

The linear-\dm\ flux model also provides a reasonable description of
the individual spectra in our sample.  For example, we show three
models with $\Delta m_{15} (B) = 0.86$, 1.12, and 1.58~mag,
respectively, in Figure~\ref{f:spec_model}.  We compare these models
to SN~2012cg, SN~2013dy, and ASASSN-14lp ($\Delta m_{15} (B) = 0.86$,
0.92, and 0.80~mag, respectively); SN~2011by and SN~2011fe ($\Delta
m_{15} (B) = 1.14$ and 1.10~mag, respectively); and SN~1992A and
SN~2011iv ($\Delta m_{15} (B) = 1.47$ and 1.69~mag, respectively).  In
general, the spectra from SNe with similar light-curve shapes have
similar spectra, and the model is also similar.  In particular,
SNe~1992A and 2011iv are remarkably similar even though they have the
largest \dm\ difference of the comparison SNe in any particular group.
Alternatively, SNe~2011by and 2011fe are the most different of any two
SNe in a group despite having the smallest \dm\ difference (and being
consistent within the uncertainties).

We can therefore use the model spectra to determine how similar a
particular spectrum is to a typical SN with its light-curve shape.
For instance, we can conclude that SN~2011fe has an excess of UV flux
relative to the typical $\Delta m_{15} (B) = 1.1$~mag SN, similar to
what one would expect from direct comparisons to SN~2011by
\citep{Foley13:met, Graham15}.  Perhaps more interesting in that
particular case is that SN~2011by has a UV spectrum more similar to
the typical SN~Ia with its light-curve shape despite its possibly
lower-than-typical luminosity \citep{Foley13:met}.

Since the UV excess for SN~2011fe at $\lambda < 2500$~\AA\ has been
interpreted as a difference in progenitor metallicity
\citep{Foley13:met, Mazzali14}, comparing a given SN~Ia
near-maximum-light UV spectrum to the model spectrum could be a useful
tool for determining the progenitor metallicity for SNe~Ia that have
no optical ``twin'' counterpart.  While our current model will be
biased by the progenitor metallicity distribution of the SN~Ia UV
sample, we can refit the model excluding any given SN, and then
compare the excluded SN to the new model.

Removing a particular SN spectrum from the sample and producing a new
model, we can examine the deviation of a particular spectrum from the
expected spectrum given its light-curve shape.  To find SNe with
potentially abnormal progenitor metallicities, one can examine the
deviation from the model spectra in the far-UV and the near-UV, where
the \citet{Lentz00} models suggest that spectra of SNe with only
progenitor metallicity differences should and should not
(respectively) differ from each other.  Examining the regions $1700 <
\lambda < 2500$~\AA\ and $2700 < \lambda < 4200$~\AA, we find two SNe
that have a median absolute fractional difference of $>$35\% for the
former and $<$25\% for the latter: SN~2011fe and ASASSN-14lp.

SN~2011fe has a spectrum consistent with that of the model in the
near-UV (median absolute fractional deviation of 8\%), but
inconsistent in the far-UV (median absolute fractional deviation of
75\%), with the far-UV flux being above the model (see
Figure~\ref{f:spec_model}).  ASASSN-14lp is also consistent with the
model in the near-UV (median absolute fractional deviation of 14\%),
but has relatively low far-UV flux (median absolute fractional
deviation of 40\%; see Figure~\ref{f:spec_model}).  Given that the
relatively high far-UV flux of SN~2011fe is interpreted as being
caused by a subsolar progenitor metallicity \citep{Foley13:met,
  Mazzali14, Baron15}, one may extrapolate to say that ASASSN-14lp has
a high progenitor metallicity.

However, we caution that the conclusions regarding the progenitor
metallicity of ASASSN-14lp rely on the other SNe in the sample with
similar light-curve shapes (namely SNe~2012cg and 2013dy).  Therefore,
a stronger conclusion is that ASASSN-14lp had a higher progenitor
metallicity than SNe~2012cg and 2013dy, without making comparisons to
SNe having significantly different light-curve shapes.

\subsection{UV Diversity}

Theory suggests that the diversity in SN~Ia UV spectra is indicative
of varying progenitor properties and explosion mechanisms.  Any
complete theory of SN~Ia explosions must explain both the general UV
spectral properties {\it and} the variance.  Similarly, knowing the UV
variance is critical for determining the cosmological utility of
rest-frame UV data for SN cosmology.

\begin{figure}
\begin{center}
\includegraphics[angle=0,width=3.2in]{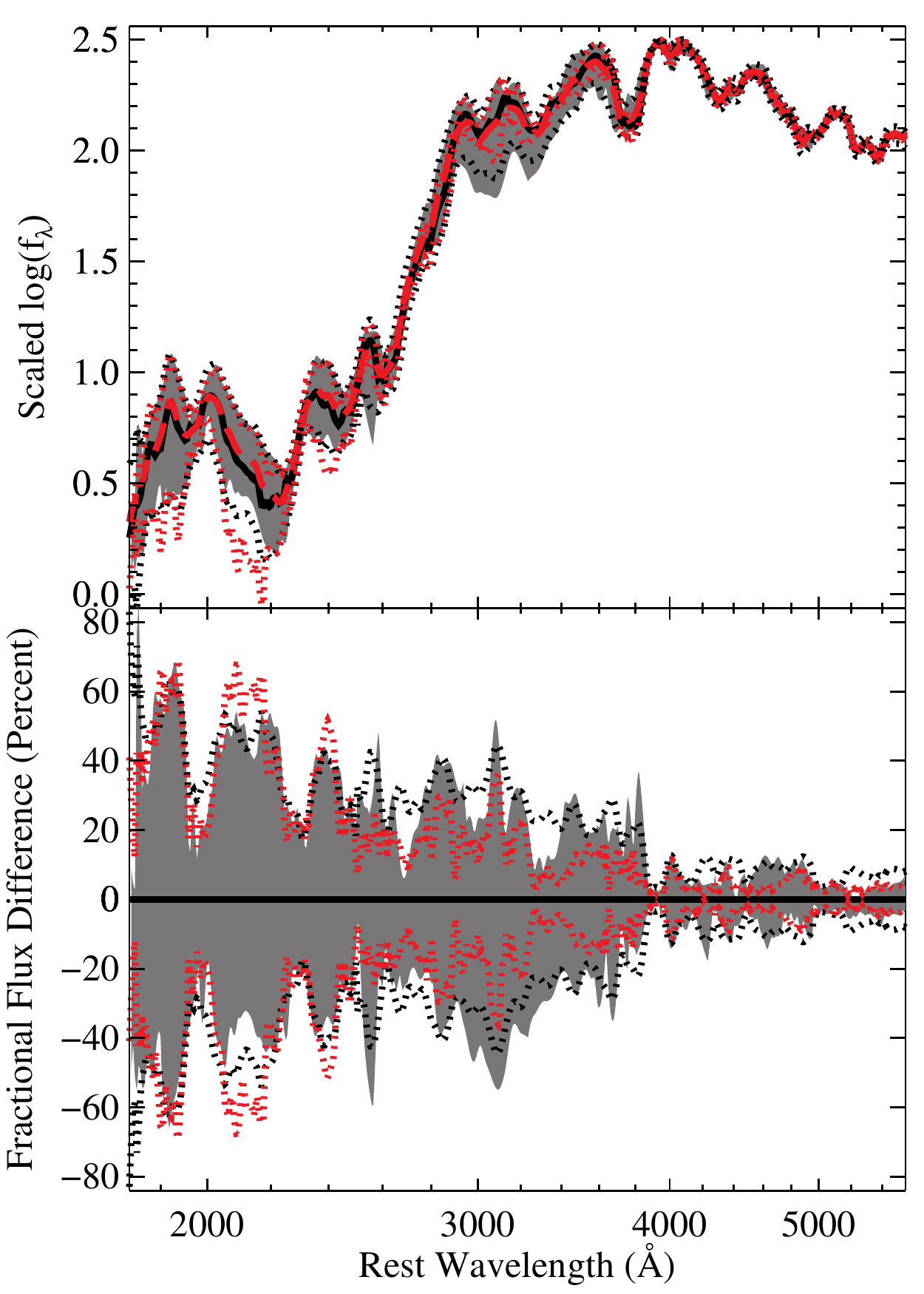}
\caption{(Top panel): Mean (black solid curve) and median (red dashed
  curve) spectra for the near-maximum-light SN~Ia spectral sample.
  The grey region represents the range of spectra for the middle
  68.27\% of spectra generated from the sample using a bootstrap
  sampling (with replacement) method.  Also plotted are the spectral
  ranges as determined by the standard deviation (black dotted curves)
  and median absolute deviation (red dotted curves).  (Bottom panel):
  Fractional flux difference for the different spectral ranges
  relative to either the mean spectrum (for the bootstrap sampling and
  standard deviation) or the median spectrum (for the median absolute
  deviation).}\label{f:boot}
\end{center}
\end{figure}

To determine the variance as a function of wavelength, we take three
similar approaches.  First, we simply measure the mean and standard
deviation of the sample.  Second, we measure the median and median
absolute deviation (MAD) of the sample.  Finally, we produce a sample
of average spectra using a bootstrap sampling (with replacement)
method \citep[see][and references therein]{Foley08:comp}.  Because the
spectra are normalized at \about 4000~\AA, the resultant spectra (and
variances) are indicative of relative spectral features, including
spectral slopes, but do not indicate differences in the overall
luminosity.  The resulting spectra are presented in
Figure~\ref{f:boot}.

Importantly, the mean and median spectra are similar at all
wavelengths, indicating that no single spectrum dramatically alters
the results.  The spectra are normalized at \about 4000~\AA, and so
the variance is naturally small there (3--8\% depending on method).
This choice does not affect the results; notably, the variance is
similarly small at \about 5000 and \about 6000~\AA\ (again 3--8\%).
However, the variance at \about 3000~\AA\ is significantly higher
(16--32\%), indicating that the exact wavelength of normalization does
not change our results.

In addition to the large increase in variance from the optical to the
near-UV, the scatter continues to increase to the far-UV.  In
particular, the median variance in the range 2500--3000~\AA\ is
roughly 24\%, while the median variance at 1800--2300~\AA\ is \about
45\%.  The increase in variance with decreasing wavelength extends the
trend seen in previous studies \citep{Foley08:comp, Foley12:sdss,
  Ellis08, Maguire12} further into the UV.


\section{Discussion and Conclusions}\label{s:disc}

We have compiled and analyzed the first sample of near-maximum-light
UV SN~Ia spectra.  This sample spans most of the parameter space of
all SNe~Ia, but still lacks examples of the fastest decliners and
those with the highest ejecta velocity.

With this sample, we examine trends between UV spectral behaviour and
optical light-curve shape, ejecta velocity, and host-galaxy
morphology.  We find that the UV spectral continuum is driven
primarily by light-curve shape, detecting only secondary correlations
with ejecta velocity or host-galaxy morphology.  We note that the lack
of high-velocity SNe~Ia in our sample prevents a thorough
investigation of its impact on spectra.  None the less, the velocities
of UV spectral features broadly correlate with those of optical
features.

Motivated by our initial investigations, we generated a UV spectral
model that depends only on light-curve shape.  This model is generally
effective at describing the behaviour of the spectra in our sample.
There are, however, notable differences.  In particular, we find that
SN~2011fe has significant excess flux at $\lambda < 2500$~\AA,
consistent with previous findings \citep{Foley13:met}.  We interpret
this difference as being caused by a particularly low-metallicity
progenitor star for SN~2011fe.

We also find that ASASSN-14lp has a deficit of flux at $\lambda <
2500$~\AA, indicating that it had a high-metallicity progenitor (at
least relative to the other SNe~Ia in our sample having similar
light-curve shapes).  This is a particularly intriguing result since
the host galaxy of ASASSN-14lp, NGC~4666, is a superwind galaxy
\citep{Dahlem97} with a high star-formation rate and likely high
metallicity.

The spectral model should allow for future spectral comparisons even
when there is no SN having a similar light-curve shape.  Since
comparing a SN spectrum to the model removes spectral differences
related to light-curve shape, any remaining differences are likely
related to other parameters such as kinetic energy, asymmetries, and
metallicity.  As our sample expands the SN parameter space (especially
to include higher-velocity SNe) and observes SNe with similar optical
properties, we will be able to better determine when a SN~Ia has
abnormal UV spectra for its light-curve shape.

Using several techniques, we found that the near-maximum-light
spectral variance increases with decreasing wavelength from \about
4000~\AA\ to at least 1800~\AA.  In particular, we find an increase in
the variance from \about 5\% to \about 25\% to \about 45\% at 4000 to
3000 to 2000~\AA, respectively. This dramatic rise in diversity at
shorter wavelengths indicates that UV observations will be of limited
utility for cosmology unless further standardization is identified.

With these results, we are now capable of discerning what
``additional'' physics contributes to the diversity of SNe~Ia beyond
the amount of $^{56}$Ni generated in the explosion.  Future UV
spectroscopy of SNe~Ia with {\it HST} and {\it Swift}, while such
capabilities exist, will fulfill this long-term goal.

\section*{Acknowledgements}



Based on observations made with the NASA/ESA Hubble Space Telescope,
obtained from the Data Archive at the Space Telescope Science
Institute, which is operated by the Association of Universities for
Research in Astronomy, Inc., under NASA contract NAS 5--26555. These
observations are associated with programs GO--4016, GO--12298,
GO--12582, GO--12592, GO--13286, and GO--13646.  We thank the STScI
staff for accommodating our target-of-opportunity programs.  A.\
Armstrong, R.\ Bohlin, S.\ Holland, S.\ Meyett, D.\ Sahnow, P.\
Sonnentrucker, and D.\ Taylor were critical for the execution of these
programs.

{\it Swift} spectroscopic observations were performed under program
GI--5080130; we are very grateful to N.\ Gehrels and the {\it Swift}
team for executing the program quickly.

R.J.F.\ gratefully acknowledges support from NASA grant 14-WPS14-0048,
NSF grant AST-1518052, and the Alfred P.\ Sloan Foundation.
G.P.\ is supported by the Ministry of Economy, Development, and
Tourism's Millennium Science Initiative through grant IC12009, awarded
to The Millennium Institute of Astrophysics, MAS.
A.V.F.\ is grateful for financial assistance from NSF grant
AST-1211916, the TABASGO Foundation, and the Christopher R. Redlich
Fund.
M.S.\ acknowledges generous support provided by the Danish Agency for
Science and Technology and Innovation realized through a Sapere Aude
Level 2 grant.

We thank the many amateur and professional astronomers who continue to
discover nearby, incredibly scientifically useful SNe {\it and}
publicly announce their discovery.
This research has made use of the NASA/IPAC Extragalactic Database
(NED) which is operated by the Jet Propulsion Laboratory, California
Institute of Technology, under contract with NASA.


\bibliographystyle{mnras}
\bibliography{../astro_refs}

\begin{thebibliography}{}
\makeatletter
\relax
\def\mn@urlcharsother{\let\do\@makeother \do\$\do\&\do\#\do\^\do\_\do\%\do\~}
\def\mn@doi{\begingroup\mn@urlcharsother \@ifnextchar [ {\mn@doi@}
  {\mn@doi@[]}}
\def\mn@doi@[#1]#2{\def\@tempa{#1}\ifx\@tempa\@empty \href
  {http://dx.doi.org/#2} {doi:#2}\else \href {http://dx.doi.org/#2} {#1}\fi
  \endgroup}
\def\mn@eprint#1#2{\mn@eprint@#1:#2::\@nil}
\def\mn@eprint@arXiv#1{\href {http://arxiv.org/abs/#1} {{\tt arXiv:#1}}}
\def\mn@eprint@dblp#1{\href {http://dblp.uni-trier.de/rec/bibtex/#1.xml}
  {dblp:#1}}
\def\mn@eprint@#1:#2:#3:#4\@nil{\def\@tempa {#1}\def\@tempb {#2}\def\@tempc
  {#3}\ifx \@tempc \@empty \let \@tempc \@tempb \let \@tempb \@tempa \fi \ifx
  \@tempb \@empty \def\@tempb {arXiv}\fi \@ifundefined
  {mn@eprint@\@tempb}{\@tempb:\@tempc}{\expandafter \expandafter \csname
  mn@eprint@\@tempb\endcsname \expandafter{\@tempc}}}

\bibitem[\protect\citeauthoryear{{Amanullah} et~al.,}{{Amanullah}
  et~al.}{2014}]{Amanullah14}
{Amanullah} R.,  et~al., 2014, \mn@doi [\apjl] {10.1088/2041-8205/788/2/L21},
  \href {http://adsabs.harvard.edu/abs/2014ApJ...788L..21A} {788, L21}

\bibitem[\protect\citeauthoryear{{Amanullah} et~al.,}{{Amanullah}
  et~al.}{2015}]{Amanullah15}
{Amanullah} R.,  et~al., 2015, \mn@doi [\mnras] {10.1093/mnras/stv1505}, \href
  {http://adsabs.harvard.edu/abs/2015MNRAS.453.3300A} {453, 3300}

\bibitem[\protect\citeauthoryear{{Baron}, {Hauschildt}, {Nugent}  \&
  {Branch}}{{Baron} et~al.}{1996}]{Baron96}
{Baron} E.,  {Hauschildt} P.~H.,  {Nugent} P.,   {Branch} D.,  1996, \mn@doi
  [\mnras] {10.1093/mnras/283.1.297}, \href
  {http://adsabs.harvard.edu/abs/1996MNRAS.283..297B} {283, 297}

\bibitem[\protect\citeauthoryear{{Baron} et~al.,}{{Baron}
  et~al.}{2015}]{Baron15}
{Baron} E.,  et~al., 2015, \mn@doi [\mnras] {10.1093/mnras/stv1951}, \href
  {http://adsabs.harvard.edu/abs/2015MNRAS.454.2549B} {454, 2549}

\bibitem[\protect\citeauthoryear{{Betoule} et~al.,}{{Betoule}
  et~al.}{2014}]{Betoule14}
{Betoule} M.,  et~al., 2014, \mn@doi [\aap] {10.1051/0004-6361/201423413},
  \href {http://adsabs.harvard.edu/abs/2014A%26A...568A..22B} {568, A22}

\bibitem[\protect\citeauthoryear{{Blondin} et~al.,}{{Blondin}
  et~al.}{2006}]{Blondin06}
{Blondin} S.,  et~al., 2006, \mn@doi [\aj] {10.1086/498724}, \href
  {http://adsabs.harvard.edu/abs/2006AJ....131.1648B} {131, 1648}

\bibitem[\protect\citeauthoryear{{Branch} et~al.,}{{Branch}
  et~al.}{2006}]{Branch06}
{Branch} D.,  et~al., 2006, \mn@doi [\pasp] {10.1086/502778}, \href
  {http://adsabs.harvard.edu/abs/2006PASP..118..560B} {118, 560}

\bibitem[\protect\citeauthoryear{{Brown} et~al.,}{{Brown}
  et~al.}{2015}]{Brown15}
{Brown} P.~J.,  et~al., 2015, \mn@doi [\apj] {10.1088/0004-637X/805/1/74},
  \href {http://adsabs.harvard.edu/abs/2015ApJ...805...74B} {805, 74}

\bibitem[\protect\citeauthoryear{{Cooke} et~al.,}{{Cooke}
  et~al.}{2011}]{Cooke11}
{Cooke} J.,  et~al., 2011, \mn@doi [\apjl] {10.1088/2041-8205/727/2/L35}, \href
  {http://adsabs.harvard.edu/abs/2011ApJ...727L..35C} {727, L35+}

\bibitem[\protect\citeauthoryear{{Dahlem}, {Petr}, {Lehnert}, {Heckman}  \&
  {Ehle}}{{Dahlem} et~al.}{1997}]{Dahlem97}
{Dahlem} M.,  {Petr} M.~G.,  {Lehnert} M.~D.,  {Heckman} T.~M.,   {Ehle} M.,
  1997, \aap, \href {http://adsabs.harvard.edu/abs/1997A%26A...320..731D} {320,
  731}

\bibitem[\protect\citeauthoryear{{Ellis} et~al.,}{{Ellis}
  et~al.}{2008}]{Ellis08}
{Ellis} R.~S.,  et~al., 2008, \mn@doi [\apj] {10.1086/524981}, \href
  {http://adsabs.harvard.edu/abs/2008ApJ...674...51E} {674, 51}

\bibitem[\protect\citeauthoryear{{Filippenko} et~al.,}{{Filippenko}
  et~al.}{1992a}]{Filippenko92:91bg}
{Filippenko} A.~V.,  et~al., 1992a, \aj, \href
  {http://adsabs.harvard.edu/cgi-bin/nph-bib_query?bibcode=1992AJ....104.1543F&db_key=AST}
  {104, 1543}

\bibitem[\protect\citeauthoryear{{Filippenko} et~al.,}{{Filippenko}
  et~al.}{1992b}]{Filippenko92:91T}
{Filippenko} A.~V.,  et~al., 1992b, \apjl, \href
  {http://adsabs.harvard.edu/cgi-bin/nph-bib_query?bibcode=1992ApJ...384L..15F&db_key=AST}
  {384, L15}

\bibitem[\protect\citeauthoryear{{Foley}}{{Foley}}{2013}]{Foley13:ca}
{Foley} R.~J.,  2013, \mn@doi [\mnras] {10.1093/mnras/stt1292}, \href
  {http://adsabs.harvard.edu/abs/2013MNRAS.435..273F} {435, 273}

\bibitem[\protect\citeauthoryear{{Foley}}{{Foley}}{2014}]{Foley14:14lp}
{Foley} R.~J.,  2014, The Astronomer's Telegram, \href
  {http://adsabs.harvard.edu/abs/2014ATel.6815....1F} {6815}

\bibitem[\protect\citeauthoryear{{Foley}}{{Foley}}{2015}]{Foley15:15f}
{Foley} R.~J.,  2015, The Astronomer's Telegram, \href
  {http://adsabs.harvard.edu/abs/2015ATel.7220....1F} {7220}

\bibitem[\protect\citeauthoryear{{Foley} \& {Kasen}}{{Foley} \&
  {Kasen}}{2011}]{Foley11:vel}
{Foley} R.~J.,  {Kasen} D.,  2011, \mn@doi [\apj] {10.1088/0004-637X/729/1/55},
  \href {http://adsabs.harvard.edu/abs/2011ApJ...729...55F} {729, 55}

\bibitem[\protect\citeauthoryear{{Foley} \& {Kirshner}}{{Foley} \&
  {Kirshner}}{2013}]{Foley13:met}
{Foley} R.~J.,  {Kirshner} R.~P.,  2013, \mn@doi [\apjl]
  {10.1088/2041-8205/769/1/L1}, \href
  {http://adsabs.harvard.edu/abs/2013ApJ...769L...1F} {769, L1}

\bibitem[\protect\citeauthoryear{{Foley} et~al.,}{{Foley}
  et~al.}{2008a}]{Foley08:comp}
{Foley} R.~J.,  et~al., 2008a, \mn@doi [\apj] {10.1086/589612}, \href
  {http://adsabs.harvard.edu/abs/2008ApJ...684...68F} {684, 68}

\bibitem[\protect\citeauthoryear{{Foley}, {Filippenko}  \& {Jha}}{{Foley}
  et~al.}{2008b}]{Foley08:uv}
{Foley} R.~J.,  {Filippenko} A.~V.,   {Jha} S.~W.,  2008b, \mn@doi [\apj]
  {10.1086/590467}, \href {http://adsabs.harvard.edu/abs/2008ApJ...686..117F}
  {686, 117}

\bibitem[\protect\citeauthoryear{{Foley}, {Sanders}  \& {Kirshner}}{{Foley}
  et~al.}{2011}]{Foley11:vgrad}
{Foley} R.~J.,  {Sanders} N.~E.,   {Kirshner} R.~P.,  2011, \mn@doi [\apj]
  {10.1088/0004-637X/742/2/89}, \href
  {http://adsabs.harvard.edu/abs/2011ApJ...742...89F} {742, 89}

\bibitem[\protect\citeauthoryear{{Foley} et~al.,}{{Foley}
  et~al.}{2012a}]{Foley12:sdss}
{Foley} R.~J.,  et~al., 2012a, \mn@doi [\aj] {10.1088/0004-6256/143/5/113},
  \href {http://adsabs.harvard.edu/abs/2012AJ....143..113F} {143, 113}

\bibitem[\protect\citeauthoryear{{Foley} et~al.,}{{Foley}
  et~al.}{2012b}]{Foley12:09ig}
{Foley} R.~J.,  et~al., 2012b, \mn@doi [\apj] {10.1088/0004-637X/744/1/38},
  \href {http://adsabs.harvard.edu/abs/2012ApJ...744...38F} {744, 38}

\bibitem[\protect\citeauthoryear{{Foley} et~al.,}{{Foley}
  et~al.}{2012c}]{Foley12:11iv}
{Foley} R.~J.,  et~al., 2012c, \mn@doi [\apjl] {10.1088/2041-8205/753/1/L5},
  \href {http://adsabs.harvard.edu/abs/2012ApJ...753L...5F} {753, L5}

\bibitem[\protect\citeauthoryear{{Foley} et~al.,}{{Foley}
  et~al.}{2014}]{Foley14:14j}
{Foley} R.~J.,  et~al., 2014, \mn@doi [\mnras] {10.1093/mnras/stu1378}, \href
  {http://adsabs.harvard.edu/abs/2014MNRAS.443.2887F} {443, 2887}

\bibitem[\protect\citeauthoryear{{Fraser} et~al.,}{{Fraser}
  et~al.}{2015}]{Fraser15}
{Fraser} M.,  et~al., 2015, The Astronomer's Telegram, \href
  {http://adsabs.harvard.edu/abs/2015ATel.7209....1F} {7209}

\bibitem[\protect\citeauthoryear{{Goobar} et~al.,}{{Goobar}
  et~al.}{2014}]{Goobar14}
{Goobar} A.,  et~al., 2014, \mn@doi [\apjl] {10.1088/2041-8205/784/1/L12},
  \href {http://adsabs.harvard.edu/abs/2014ApJ...784L..12G} {784, L12}

\bibitem[\protect\citeauthoryear{{Graham} et~al.,}{{Graham}
  et~al.}{2015}]{Graham15}
{Graham} M.~L.,  et~al., 2015, \mn@doi [\mnras] {10.1093/mnras/stu2221}, \href
  {http://adsabs.harvard.edu/abs/2015MNRAS.446.2073G} {446, 2073}

\bibitem[\protect\citeauthoryear{{Hamuy}, {Trager}, {Pinto}, {Phillips},
  {Schommer}, {Ivanov}  \& {Suntzeff}}{{Hamuy} et~al.}{2000}]{Hamuy00}
{Hamuy} M.,  {Trager} S.~C.,  {Pinto} P.~A.,  {Phillips} M.~M.,  {Schommer}
  R.~A.,  {Ivanov} V.,   {Suntzeff} N.~B.,  2000, \mn@doi [\aj]
  {10.1086/301527}, \href {http://adsabs.harvard.edu/abs/2000AJ....120.1479H}
  {120, 1479}

\bibitem[\protect\citeauthoryear{{Hillebrandt} \& {Niemeyer}}{{Hillebrandt} \&
  {Niemeyer}}{2000}]{Hillebrandt00}
{Hillebrandt} W.,  {Niemeyer} J.~C.,  2000, \mn@doi [\araa]
  {10.1146/annurev.astro.38.1.191}, \href
  {http://adsabs.harvard.edu/abs/2000ARA%26A..38..191H} {38, 191}

\bibitem[\protect\citeauthoryear{{H\"{o}flich}, {Wheeler}  \&
  {Thielemann}}{{H\"{o}flich} et~al.}{1998}]{Hoflich98}
{H\"{o}flich} P.,  {Wheeler} J.~C.,   {Thielemann} F.-K.,  1998, \mn@doi [\apj]
  {10.1086/305327}, \href {http://adsabs.harvard.edu/abs/1998ApJ...495..617H}
  {495, 617}

\bibitem[\protect\citeauthoryear{{Holoien} et~al.,}{{Holoien}
  et~al.}{2014}]{Holoien14:14lp}
{Holoien} T.~W.-S.,  et~al., 2014, The Astronomer's Telegram, \href
  {http://adsabs.harvard.edu/abs/2014ATel.6795....1H} {6795}

\bibitem[\protect\citeauthoryear{{Howell}}{{Howell}}{2001}]{Howell01}
{Howell} D.~A.,  2001, \mn@doi [\apjl] {10.1086/321702}, \href
  {http://adsabs.harvard.edu/abs/2001ApJ...554L.193H} {554, L193}

\bibitem[\protect\citeauthoryear{{Im}, {Choi}, {Yoon}, {Kim}, {Ehgamberdiev},
  {Monard}  \& {Sung}}{{Im} et~al.}{2015}]{Im15}
{Im} M.,  {Choi} C.,  {Yoon} S.-C.,  {Kim} J.-W.,  {Ehgamberdiev} S.~A.,
  {Monard} L.~A.~G.,   {Sung} H.-I.,  2015, \mn@doi [\apjs]
  {10.1088/0067-0049/221/1/22}, \href
  {http://adsabs.harvard.edu/abs/2015ApJS..221...22I} {221, 22}

\bibitem[\protect\citeauthoryear{{Kasen} \& {Woosley}}{{Kasen} \&
  {Woosley}}{2007}]{Kasen07:wlr}
{Kasen} D.,  {Woosley} S.~E.,  2007, \mn@doi [\apj] {10.1086/510375}, \href
  {http://adsabs.harvard.edu/abs/2007ApJ...656..661K} {656, 661}

\bibitem[\protect\citeauthoryear{{Kelly}, {Hicken}, {Burke}, {Mandel}  \&
  {Kirshner}}{{Kelly} et~al.}{2010}]{Kelly10}
{Kelly} P.~L.,  {Hicken} M.,  {Burke} D.~L.,  {Mandel} K.~S.,   {Kirshner}
  R.~P.,  2010, \mn@doi [\apj] {10.1088/0004-637X/715/2/743}, \href
  {http://adsabs.harvard.edu/abs/2010ApJ...715..743K} {715, 743}

\bibitem[\protect\citeauthoryear{{Kim} et~al.,}{{Kim} et~al.}{2014}]{Kim14}
{Kim} A.~G.,  et~al., 2014, \mn@doi [\apj] {10.1088/0004-637X/784/1/51}, \href
  {http://adsabs.harvard.edu/abs/2014ApJ...784...51K} {784, 51}

\bibitem[\protect\citeauthoryear{{Kirshner} et~al.,}{{Kirshner}
  et~al.}{1993}]{Kirshner93}
{Kirshner} R.~P.,  et~al., 1993, \mn@doi [\apj] {10.1086/173188}, \href
  {http://adsabs.harvard.edu/abs/1993ApJ...415..589K} {415, 589}

\bibitem[\protect\citeauthoryear{{Lampeitl} et~al.,}{{Lampeitl}
  et~al.}{2010}]{Lampeitl10:host}
{Lampeitl} H.,  et~al., 2010, \mn@doi [\apj] {10.1088/0004-637X/722/1/566},
  \href {http://adsabs.harvard.edu/abs/2010ApJ...722..566L} {722, 566}

\bibitem[\protect\citeauthoryear{{Leibundgut} et~al.,}{{Leibundgut}
  et~al.}{1993}]{Leibundgut93}
{Leibundgut} B.,  et~al., 1993, \mn@doi [\aj] {10.1086/116427}, \href
  {http://adsabs.harvard.edu/abs/1993AJ....105..301L} {105, 301}

\bibitem[\protect\citeauthoryear{{Lentz}, {Baron}, {Branch}, {Hauschildt}  \&
  {Nugent}}{{Lentz} et~al.}{2000}]{Lentz00}
{Lentz} E.~J.,  {Baron} E.,  {Branch} D.,  {Hauschildt} P.~H.,   {Nugent}
  P.~E.,  2000, \mn@doi [\apj] {10.1086/308400}, \href
  {http://adsabs.harvard.edu/abs/2000ApJ...530..966L} {530, 966}

\bibitem[\protect\citeauthoryear{{Lentz}, {Baron}, {Branch}  \&
  {Hauschildt}}{{Lentz} et~al.}{2001}]{Lentz01}
{Lentz} E.~J.,  {Baron} E.,  {Branch} D.,   {Hauschildt} P.~H.,  2001, \mn@doi
  [\apj] {10.1086/322239}, \href
  {http://adsabs.harvard.edu/abs/2001ApJ...557..266L} {557, 266}

\bibitem[\protect\citeauthoryear{{Maguire} et~al.,}{{Maguire}
  et~al.}{2012}]{Maguire12}
{Maguire} K.,  et~al., 2012, \mn@doi [\mnras]
  {10.1111/j.1365-2966.2012.21909.x}, \href
  {http://adsabs.harvard.edu/abs/2012MNRAS.426.2359M} {426, 2359}

\bibitem[\protect\citeauthoryear{{Mazzali} \& {Podsiadlowski}}{{Mazzali} \&
  {Podsiadlowski}}{2006}]{Mazzali06}
{Mazzali} P.~A.,  {Podsiadlowski} P.,  2006, \mn@doi [\mnras]
  {10.1111/j.1745-3933.2006.00165.x}, \href
  {http://adsabs.harvard.edu/abs/2006MNRAS.369L..19M} {369, L19}

\bibitem[\protect\citeauthoryear{{Mazzali}, {Nomoto}, {Cappellaro}, {Nakamura},
  {Umeda}  \& {Iwamoto}}{{Mazzali} et~al.}{2001}]{Mazzali01}
{Mazzali} P.~A.,  {Nomoto} K.,  {Cappellaro} E.,  {Nakamura} T.,  {Umeda} H.,
  {Iwamoto} K.,  2001, \mn@doi [\apj] {10.1086/318428}, \href
  {http://adsabs.harvard.edu/abs/2001ApJ...547..988M} {547, 988}

\bibitem[\protect\citeauthoryear{{Mazzali} et~al.,}{{Mazzali}
  et~al.}{2014}]{Mazzali14}
{Mazzali} P.~A.,  et~al., 2014, \mn@doi [\mnras] {10.1093/mnras/stu077}, \href
  {http://adsabs.harvard.edu/abs/2014MNRAS.439.1959M} {439, 1959}

\bibitem[\protect\citeauthoryear{{Monard} et~al.,}{{Monard}
  et~al.}{2015}]{Monard15}
{Monard} L.~A.~G.,  et~al., 2015, Central Bureau Electronic Telegrams, \href
  {http://adsabs.harvard.edu/abs/2015CBET.4081....1M} {4081}

\bibitem[\protect\citeauthoryear{{Nugent}, {Phillips}, {Baron}, {Branch}  \&
  {Hauschildt}}{{Nugent} et~al.}{1995}]{Nugent95}
{Nugent} P.,  {Phillips} M.,  {Baron} E.,  {Branch} D.,   {Hauschildt} P.,
  1995, \mn@doi [\apjl] {10.1086/309846}, \href
  {http://adsabs.harvard.edu/abs/1995ApJ...455L.147N} {455, L147}

\bibitem[\protect\citeauthoryear{{Nugent}, {Kim}  \& {Perlmutter}}{{Nugent}
  et~al.}{2002}]{Nugent02}
{Nugent} P.,  {Kim} A.,   {Perlmutter} S.,  2002, \pasp, \href
  {http://adsabs.harvard.edu/abs/2002PASP..114..803N} {114, 803}

\bibitem[\protect\citeauthoryear{{Pan} et~al.,}{{Pan}
  et~al.}{2015}]{Pan15:13dy}
{Pan} Y.-C.,  et~al., 2015, \mn@doi [\mnras] {10.1093/mnras/stv1605}, \href
  {http://adsabs.harvard.edu/abs/2015MNRAS.452.4307P} {452, 4307}

\bibitem[\protect\citeauthoryear{{Pereira} et~al.,}{{Pereira}
  et~al.}{2013}]{Pereira13}
{Pereira} R.,  et~al., 2013, \mn@doi [\aap] {10.1051/0004-6361/201221008},
  \href {http://adsabs.harvard.edu/abs/2013A%26A...554A..27P} {554, A27}

\bibitem[\protect\citeauthoryear{{Perlmutter} et~al.,}{{Perlmutter}
  et~al.}{1999}]{Perlmutter99}
{Perlmutter} S.,  et~al., 1999, \mn@doi [\apj] {10.1086/307221}, \href
  {http://adsabs.harvard.edu/abs/1999ApJ...517..565P} {517, 565}

\bibitem[\protect\citeauthoryear{{Phillips}}{{Phillips}}{1993}]{Phillips93}
{Phillips} M.~M.,  1993, \mn@doi [\apjl] {10.1086/186970}, \href
  {http://adsabs.harvard.edu/abs/1993ApJ...413L.105P} {413, L105}

\bibitem[\protect\citeauthoryear{{Phillips} et~al.,}{{Phillips}
  et~al.}{1987}]{Phillips87}
{Phillips} M.~M.,  et~al., 1987, \pasp, \href
  {http://adsabs.harvard.edu/abs/1987PASP...99..592P} {99, 592}

\bibitem[\protect\citeauthoryear{{Phillips}, {Wells}, {Suntzeff}, {Hamuy},
  {Leibundgut}, {Kirshner}  \& {Foltz}}{{Phillips} et~al.}{1992}]{Phillips92}
{Phillips} M.~M.,  {Wells} L.~A.,  {Suntzeff} N.~B.,  {Hamuy} M.,  {Leibundgut}
  B.,  {Kirshner} R.~P.,   {Foltz} C.~B.,  1992, \mn@doi [\aj]
  {10.1086/116177}, \href {http://adsabs.harvard.edu/abs/1992AJ....103.1632P}
  {103, 1632}

\bibitem[\protect\citeauthoryear{{Phillips}, {Lira}, {Suntzeff}, {Schommer},
  {Hamuy}  \& {Maza}}{{Phillips} et~al.}{1999}]{Phillips99}
{Phillips} M.~M.,  {Lira} P.,  {Suntzeff} N.~B.,  {Schommer} R.~A.,  {Hamuy}
  M.,   {Maza} J.,  1999, \mn@doi [\aj] {10.1086/301032}, \href
  {http://adsabs.harvard.edu/abs/1999AJ....118.1766P} {118, 1766}

\bibitem[\protect\citeauthoryear{{Pinto} \& {Eastman}}{{Pinto} \&
  {Eastman}}{2000}]{Pinto00}
{Pinto} P.~A.,  {Eastman} R.~G.,  2000, \mn@doi [\apj] {10.1086/308380}, \href
  {http://adsabs.harvard.edu/abs/2000ApJ...530..757P} {530, 757}

\bibitem[\protect\citeauthoryear{{Podsiadlowski}, {Mazzali}, {Lesaffre}, {Wolf}
   \& {Forster}}{{Podsiadlowski} et~al.}{2006}]{Podsiadlowski06}
{Podsiadlowski} P.,  {Mazzali} P.~A.,  {Lesaffre} P.,  {Wolf} C.,   {Forster}
  F.,  2006, preprint, \href
  {http://adsabs.harvard.edu/abs/2006astro.ph..8324P} {} (\mn@eprint {}
  {astro-ph/0608324})

\bibitem[\protect\citeauthoryear{{Rest} et~al.,}{{Rest} et~al.}{2014}]{Rest14}
{Rest} A.,  et~al., 2014, \mn@doi [\apj] {10.1088/0004-637X/795/1/44}, \href
  {http://adsabs.harvard.edu/abs/2014ApJ...795...44R} {795, 44}

\bibitem[\protect\citeauthoryear{{Riess} et~al.,}{{Riess}
  et~al.}{1998}]{Riess98:Lambda}
{Riess} A.~G.,  et~al., 1998, \aj, \href
  {http://adsabs.harvard.edu/cgi-bin/nph-bib_query?bibcode=1998AJ....116.1009R&db_key=AST}
  {116, 1009}

\bibitem[\protect\citeauthoryear{{Sauer} et~al.,}{{Sauer}
  et~al.}{2008}]{Sauer08}
{Sauer} D.~N.,  et~al., 2008, \mn@doi [\mnras]
  {10.1111/j.1365-2966.2008.14018.x}, \href
  {http://adsabs.harvard.edu/abs/2008MNRAS.391.1605S} {391, 1605}

\bibitem[\protect\citeauthoryear{{Schlafly} \& {Finkbeiner}}{{Schlafly} \&
  {Finkbeiner}}{2011}]{Schlafly11}
{Schlafly} E.~F.,  {Finkbeiner} D.~P.,  2011, \mn@doi [\apj]
  {10.1088/0004-637X/737/2/103}, \href
  {http://adsabs.harvard.edu/abs/2011ApJ...737..103S} {737, 103}

\bibitem[\protect\citeauthoryear{{Schlegel}, {Finkbeiner}  \&
  {Davis}}{{Schlegel} et~al.}{1998}]{Schlegel98}
{Schlegel} D.~J.,  {Finkbeiner} D.~P.,   {Davis} M.,  1998, \apj, \href
  {http://adsabs.harvard.edu/cgi-bin/nph-bib_query?bibcode=1998ApJ...500..525S&db_key=AST}
  {500, 525}

\bibitem[\protect\citeauthoryear{{Shappee} et~al.,}{{Shappee}
  et~al.}{2015}]{Shappee15}
{Shappee} B.~J.,  et~al., 2015, preprint, \href
  {http://adsabs.harvard.edu/abs/2015arXiv150704257S} {} (\mn@eprint {arXiv}
  {1507.04257})

\bibitem[\protect\citeauthoryear{{Silverman} et~al.,}{{Silverman}
  et~al.}{2012}]{Silverman12:12cg}
{Silverman} J.~M.,  et~al., 2012, \mn@doi [\apjl] {10.1088/2041-8205/756/1/L7},
  \href {http://adsabs.harvard.edu/abs/2012ApJ...756L...7S} {756, L7}

\bibitem[\protect\citeauthoryear{{Silverman}, {Ganeshalingam}  \&
  {Filippenko}}{{Silverman} et~al.}{2013}]{Silverman13}
{Silverman} J.~M.,  {Ganeshalingam} M.,   {Filippenko} A.~V.,  2013, \mn@doi
  [\mnras] {10.1093/mnras/sts674}, \href
  {http://adsabs.harvard.edu/abs/2013MNRAS.tmp..529S} {p.~529}

\bibitem[\protect\citeauthoryear{{Sullivan} et~al.,}{{Sullivan}
  et~al.}{2010}]{Sullivan10}
{Sullivan} M.,  et~al., 2010, \mn@doi [\mnras]
  {10.1111/j.1365-2966.2010.16731.x}, \href
  {http://adsabs.harvard.edu/abs/2010MNRAS.406..782S} {406, 782}

\bibitem[\protect\citeauthoryear{{Thorstensen} et~al.,}{{Thorstensen}
  et~al.}{2014}]{Thorstensen14}
{Thorstensen} J.,  et~al., 2014, The Astronomer's Telegram, \href
  {http://adsabs.harvard.edu/abs/2014ATel.6801....1T} {6801}

\bibitem[\protect\citeauthoryear{{Timmes}, {Brown}  \& {Truran}}{{Timmes}
  et~al.}{2003}]{Timmes03}
{Timmes} F.~X.,  {Brown} E.~F.,   {Truran} J.~W.,  2003, \mn@doi [\apjl]
  {10.1086/376721}, \href {http://adsabs.harvard.edu/abs/2003ApJ...590L..83T}
  {590, L83}

\bibitem[\protect\citeauthoryear{{Wang} et~al.,}{{Wang}
  et~al.}{2009}]{Wang09:2pop}
{Wang} X.,  et~al., 2009, \mn@doi [\apjl] {10.1088/0004-637X/699/2/L139}, \href
  {http://adsabs.harvard.edu/abs/2009ApJ...699L.139W} {699, L139}

\bibitem[\protect\citeauthoryear{{Wang} et~al.,}{{Wang}
  et~al.}{2012}]{Wang12:uv}
{Wang} X.,  et~al., 2012, \mn@doi [\apj] {10.1088/0004-637X/749/2/126}, \href
  {http://adsabs.harvard.edu/abs/2012ApJ...749..126W} {749, 126}

\makeatother
\end{thebibliography}


\end{document}